\documentclass[aps,prl,twocolumn,floatfix]{revtex4}
\usepackage{eurosym}
\usepackage{amsfonts}
\usepackage{mathrsfs}
\usepackage{color}
\usepackage{xspace}
\usepackage{subfigure}
\usepackage{longtable}
\usepackage{xspace}
\usepackage{multirow}
\usepackage{tabularx}
\usepackage{amsmath}
\usepackage{amssymb}
\usepackage{graphicx}
\usepackage{dcolumn}
\usepackage{bm}
\usepackage[colorlinks,urlcolor=blue,citecolor=blue,linkcolor=blue]{hyperref}
\usepackage[normalem]{ulem}

\newcommand{\R}{\mathbb{R}}

\begin{document}

\title{Dissipative shock waves generated by a quantum-mechanical piston}
\author{Maren E.~Mossman$^{1}$}
\author{Mark A.~Hoefer$^{2}$}
\thanks{hoefer@colorado.edu}
\author{Keith Julien$^{2}$}
\author{P.~G.~Kevrekidis$^{3}$}
\author{P.~Engels$^{1}$}
\thanks{engels@wsu.edu}

\begin{abstract}
  {\bf Abstract} 
  The piston shock problem is a prototypical example of strongly nonlinear fluid flow that enables the experimental exploration of fluid dynamics in extreme regimes.  
  Here we investigate this problem for a nominally dissipationless, superfluid Bose-Einstein condensate and observe rich dynamics including the formation of a plateau region, a non-expanding shock front, and rarefaction waves.
  Many aspects of the observed dynamics follow predictions of classical dissipative---rather than superfluid dispersive---shock theory.  
  The emergence of dissipative-like dynamics is attributed to the decay of large amplitude excitations at the shock front into turbulent vortex excitations which allow us to invoke an eddy viscosity hypothesis.
  Our experimental observations are accompanied by numerical simulations of the mean field, Gross-Pitaevskii equation that exhibit quantitative agreement with no fitting parameters.  
  This work provides an avenue for the investigation of quantum shock waves and turbulence in channel geometries, which are currently the focus of intense research efforts.
\end{abstract}

\affiliation{
$^{1}$Department of Physics and Astronomy, Washington State University, Pullman, WA, USA 99164 \\
$^{2}$ Department of Applied Mathematics, University of Colorado, Boulder, CO, USA  80309-0526\\
$^{3}$ Department of Mathematics and Statistics, University of Massachusetts, Amherst, MA, USA 01003-4515}
\maketitle

From the generation of localized solitons and quantized vortices \cite{Kevrekidis2008} to the extended coherence of dispersive shock waves \cite{El2016}, quantum hydrodynamics exhibit an intriguingly rich phenomenology. 
While many pioneering observations have been made in superfluid helium \cite{Donnelly1986,Barenghi2001}, dilute-gas Bose-Einstein condensates (BECs) provide an exceptionally versatile medium in which to access quantum hydrodynamics~\cite{Bradley2012}. 
The experimental control and theoretical understanding of BECs enable novel techniques for entering new quantum hydrodynamic regimes. 
A central focus of superfluid helium studies has been the investigation of quantum turbulence, including the origin of dissipation within the system \cite{Onsager1949}. 
Despite strong experimental and theoretical research efforts spanning many decades (see \cite{Tsubota2017} and references therein), quantum turbulence still poses many open questions. 
For example, while many aspects of quantum turbulence in a homogeneous system have been clarified, the nature of quantum turbulence in channel geometries is now under intense investigation in superfluid helium systems.  
In dilute gas BECs, quantum turbulence has been experimentally observed only in a very limited number of settings so far. 
Those include the generation of vortex turbulence in stirred BECs \cite{Henn2009,Neely2013,Kwon2014},
the observation of weak-wave turbulence in a shaken BEC confined in a box potential \cite{Navon2016}, and the observation of spinor turbulence \cite{Seo2016,Kang2017}. 
Aspects involving the definition of a superfluid Reynolds number~\cite{Reeves2015}, or energy and enstrophy cascades~\cite{Nore1997, Proment2009, Reeves2017b}, remain under very active theoretical investigation. 
A discussion of relevant experimental realizations but also of theoretical attempts to study the problem has recently been compiled in~\cite{Tsatsos2016}.

Here we introduce a setting for the observation of rich quantum hydrodynamics by studying a BEC piston shock in a channel geometry. 
The piston shock is a paradigmatic example of strongly nonlinear flow, probing hydrodynamics in an extreme regime. 
For a one-dimensional channel, the BEC piston shock is theoretically predicted to be an expanding, coherent dispersive shock wave (DSW) with rank ordered, nonlinear oscillations \cite{Hoefer2008}. 
A related setting, the collision of two BECs with strong transverse confinement in a channel, has been shown experimentally and theoretically to give rise to similar dynamics, such as the continuous transformation of a sinusoidal interference pattern into a train of dark solitons that was interpreted as two adjacent dispersive shock waves \cite{Hoefer2009}. 
However, in the presence of weaker transverse confinement, collision experiments in BECs \cite{Chang2008} (and also in dilute Fermi gases \cite{Joseph2011}) cannot be described by DSWs but by two counterpropagating, viscous or dissipative shock waves (VSWs) \cite{Lowman2013}. 
Two features that distinguish a VSW---however weak the dissipation---from a DSW are: i) its shock width is independent of time and proportional to the medium's dissipation, ii) its speed is uniquely determined by the Rankine-Hugoniot jump relations \cite{Liepmann1957}. 
In contrast, a DSW exhibits an expanding series of rank-ordered oscillations with two edge speeds, each of which satisfies DSW closure relations that are entirely different from the Rankine-Hugoniot relations \cite{El2016}.

The fully three-dimensional BEC piston shock problem explored here provides a clean setting for the quantitative study of the roles of dispersion and dissipation in nonlinear quantum hydrodynamics. 
We observe the generation of non-expanding, large-scale shocks that satisfy the Rankine-Hugoniot jump relations, image features indicative of vortex turbulence at small, healing length scales, and perform numerical simulations of the conservative, mean-field Gross-Pitaevskii (GP) equation that quantitatively agree with experiment. 
Piston compression is found to continually generate two distinct wave-field components: sound waves and solitons.  The solitons breakup into vortices via the well-known snake instability.  
We argue that the decay of large amplitude excitations generated at the shock front into vortex excitations lead to an emergent dissipative like behaviour in a coarse-grained description of the fluid. Consequently, piston compression provides both the generation mechanism for turbulence into which large amplitude soliton like excitations dissipate and the sustenance of a subsonic to supersonic shock front.
Thus, the piston shock problem also opens a pathway to the study of quantum turbulence with BECs in channel geometries.

\section{Results}

\subsection{Overview of Shock Dynamics}

\begin{figure}
\includegraphics[width=9cm]{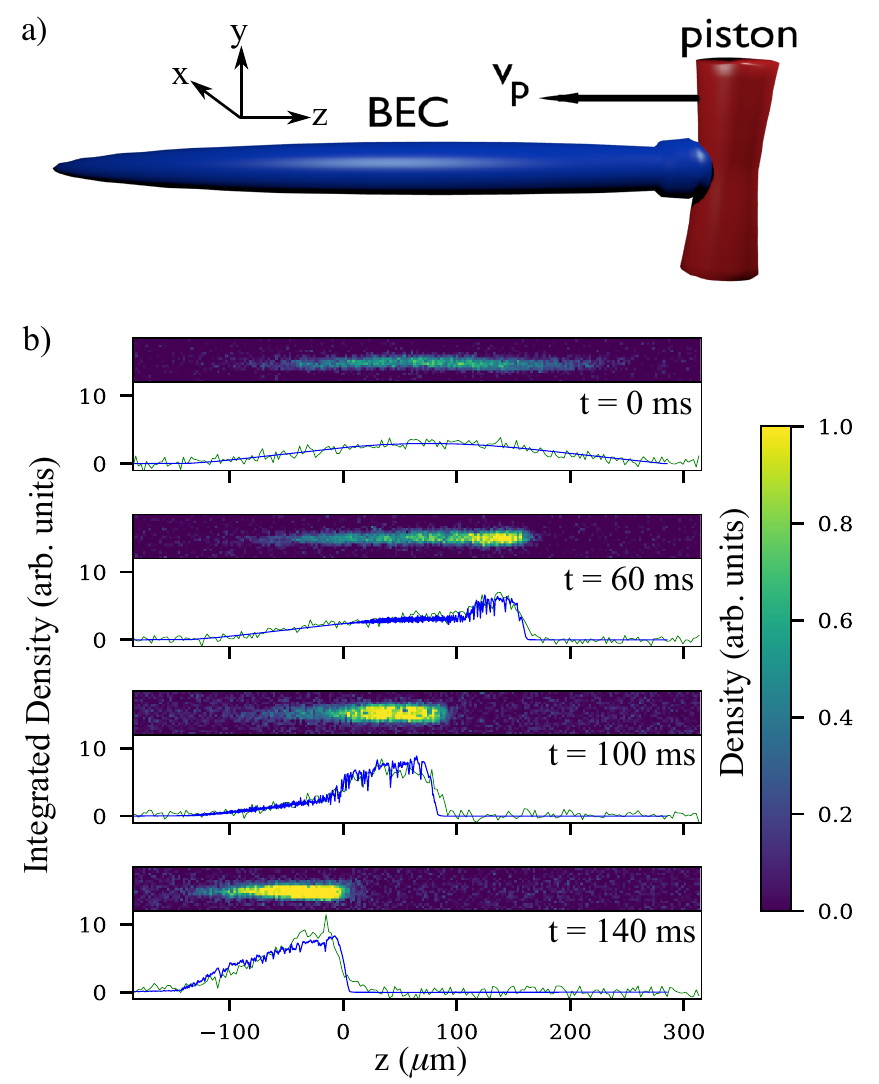}
\caption{\emph{Experimental Setup and Integrated Cross Sections.} 
\textbf{a,} A repulsive barrier (piston) is swept from right to left through a BEC with speed $v_p$.
 A bulge forms at the interface of the BEC and the piston. 
 Image is not to scale. 
 \textbf{b,} Experimental images and corresponding integrated cross sections for experiment (green) and numerics (blue) at times $t = 0, 60, 100, \text{ and } 140$~ms into a $v_p = 2$~mm s$^{-1}$ sweep.}
\label{Fig:SweepSequence}
\end{figure}

The general setting for our study is comprised of an elongated BEC confined by a cigar-shaped harmonic trap. 
A repulsive barrier, the height of which exceeds the chemical potential of the BEC by a factor of 10, is created by the dipole force of a far detuned laser beam.
Initially, the barrier is placed outside the BEC and is then moved through the BEC at a constant speed denoted $v_{\text{p}}$. 
Our axis convention is such that the weakly confined z-axis is oriented horizontally [Fig.~\ref{Fig:SweepSequence}(a)]. 
For reference, the bulk speed of sound in the center of the initial, unperturbed BEC is $c_{\text{s,bulk}} \approx 2.47~\text{mm s}^{-1}$, and the speed with which sound pulses travel along the long axis of the initial, unperturbed BEC is calculated to be $c_{\text{s,1d}} = c_{\text{s,bulk}}/\sqrt{2} \approx 1.75~\text{mm s}^{-1}$ in the Thomas-Fermi regime \cite{Zaremba1998}. 
Absorption images are taken at sequential times during the piston sweep to analyze the resulting dynamics.

For sweep speeds near or exceeding the speed of sound, $c_{\text{s,1d}}$, such as the case shown in Fig.~\ref{Fig:SweepSequence} (b), the dynamics are intriguingly rich. 
At $t = 0$~ms, the piston is located just to the right side of the BEC. 
As the piston enters the BEC, a pronounced density spike forms near the piston front.  
As the piston sweeps through the cloud, a plateau region of high density develops in front of the piston (Fig.~\ref{Fig:SweepSequence}(b), 60~ms). 
The leading (left) edge of the plateau forms a steep shock front, where the density rapidly drops from the plateau density to the initial BEC density. 
Once the plateau reaches the left wing of the BEC, the shock front deforms and approaches the shape of a rarefaction wave (see discussion below).
The onset of this behavior can be seen in Fig.~\ref{Fig:SweepSequence}(b) at $t = 140$~ms.

The experimentally observed dynamics are in excellent agreement with our three-dimensional numerical simulations based on the Gross-Pitaevskii equation (GPE) with no fitting parameters. 
Parameters utilized in the numerics are taken directly from experiment. 
For more details, see Methods.

\subsection{Characterization of the Shock Dynamics}

\begin{figure}
  \includegraphics[width=9cm]{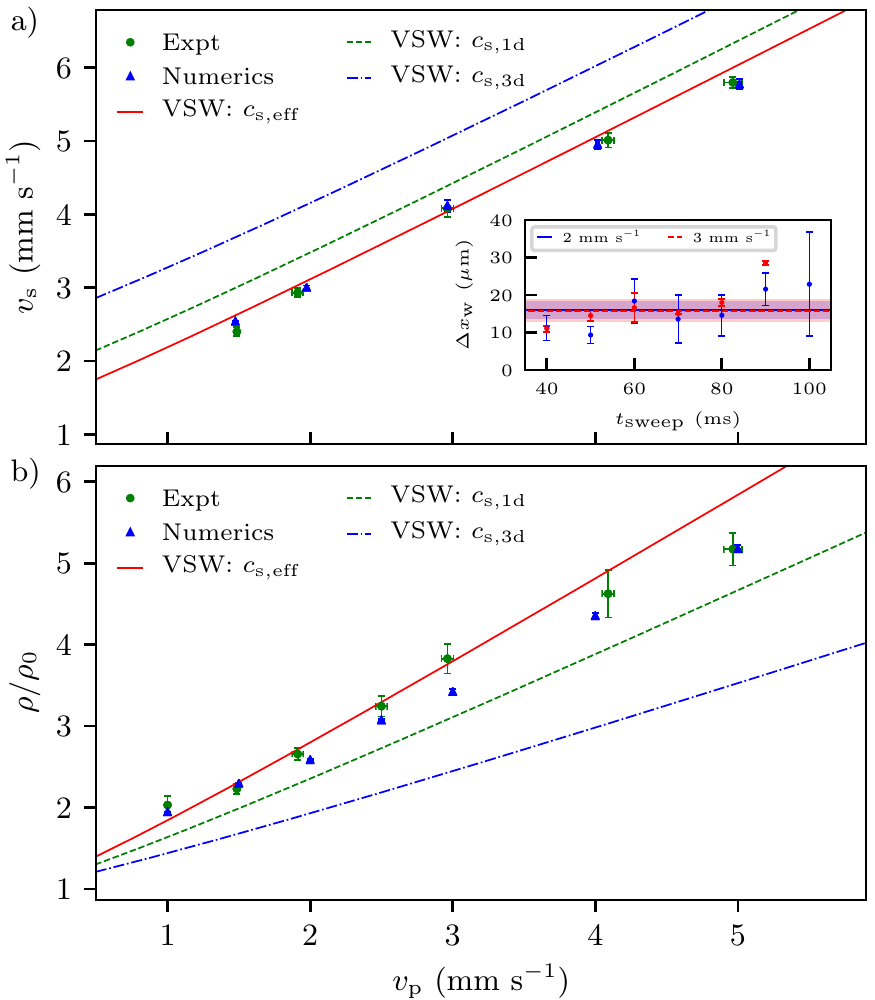}
  \caption{\emph{Shock Speed, Peak Density, and Shock Width vs.~Piston Speed.} 
  Analyzed experimental (green dots) and numerical (blue triangles) results for increasing piston velocity are plotted with overlaid theory curves for VSW (red solid line), using an effective $c_\text{s,eff} = 1.35$~mm s$^{-1}$ obtained from a fit of the experimental and numerical data to the VSW theory prediction. 
  VSW theory curves are also calculated for $c_\text{s,1d} = 1.75$~mm s$^{-1}$ (green dashed) and $c_\text{s,bulk} = 2.47$ mm s$^{-1}$ (blue dot-dashed). 
  \textbf{a,}~Shock front speed.
  \textbf{Inset,}~Shock front width at different times for piston speeds 2 mm s$^{-1}$ (blue solid) and 3 mm s$^{-1}$ (red dashed), with corresponding error (shaded regions). 
  \textbf{b,}~The normalized plateau density is determined by measuring the plateau integrated density when the piston reaches the center of the BEC. 
  Weighted mean $\pm$ s.d. are plotted for three (a) and five (b) sets of data. 
  For more information, see text.}
\label{Fig:MainResults}
\end{figure}

For a quantitative analysis of the piston shock wave dynamics, we consider three characteristic features: the shock propagation speed, the shock width, and the plateau density. 
Their behavior as a function of the piston speed is shown in Fig.~\ref{Fig:MainResults}. 
The shock propagation speed is determined by tracking the position of the shock front as it moves through the central part of the BEC where it exhibits an approximately constant speed. 
For both experimental and numerical data, the shock front edge is obtained by calculating integrated cross sections and subtracting the cross section of an unperturbed BEC in the absence of a piston sweep.
In the subtracted plots, the shock front is fit with a straight line.
The zero intercept of the fit is recorded as the position of the shock front, and its speed is the shock propagation speed.
Experiment and corresponding numerics indicate an approximately linear increase in the shock speed with increasing piston speed (Fig.~\ref{Fig:MainResults}a). 
The inset of Fig.~\ref{Fig:MainResults}a shows the evolution of the shock front width during sweeps with $v_\text{p} = 2~$mm s$^{-1}$ (blue solid) and $v_\text{p} = 3~$mm s$^{-1}$ (red dashed).
The width is determined by following the shock front as it is driven through the BEC.
Similar to finding the shock front position, we calculate integrated cross sections, subtract an unperturbed cross section and record the spacing between the zero intercept and the edge of the plateau region.
Both piston speeds result in a time-averaged width of $\Delta x_\text{w} = 15.5~\mu$m. 
This constant width over the course of a sweep is indicative of VSWs.
The plateau density, as a function of piston speed, is determined by averaging the density of the plateau region that has formed in front of the barrier when the piston reaches the center of the BEC. 
This is where the initial density of the BEC is nominally uniform. 
The observed dependence on the piston speed is approximately linear for medium piston speeds, but undershoots a linear trend at large piston speeds (Fig.~\ref{Fig:MainResults}b). 
In all cases, we see remarkable agreement between experiment and numerics based on the GPE.

Because dilute-gas BECs are typically modeled as inviscid, dispersive superfluids, an obvious approach is to compare the observed behavior with that of BEC dispersive shock wave (DSW) theory \cite{El1995,Kamchatnov2004,Hoefer2006} (see also the review \cite{El2016}).  
The one-dimensional DSW theory for the BEC piston problem was presented in \cite{Hoefer2008} (see also \cite{Kamchatnov2010}).  
A dispersive shock wave is characterized by an expanding, coherent nonlinear waveform with a trailing large amplitude soliton edge, rank-ordered interior oscillations, and a leading small amplitude, harmonically oscillatory edge.  
However, the best fit to the predicted DSW soliton edge speed and plateau height is poor (see Supplementary Fig.~1).  
On the other hand, VSW theory, where a narrow, planar shock front is assumed to propagate through a uniform, weakly viscous medium (see, for example, \cite{Leveque2002}), leads to a consistent description of both the shock speed data and the plateau height data if an effective speed of sound of $c_\text{s,eff} = 1.35$~mm s$^{-1}$ is assumed in the calculation.  
The resulting VSW predictions are shown as the red solid lines in Fig.~\ref{Fig:MainResults}.
VSW speed and plateau height are independent of viscosity, and are determined by the flow conditions in front of and behind the shock front according to the Rankine-Hugoniot jump conditions \cite{Leveque2002}.  
Relevant details of this analysis are given in Supplementary Note 1.  
We note that the obtained effective speed of sound is lower than the bulk speed of sound in the BEC center ($c_\text{s,bulk} \approx 2.47$~mm s$^{-1}$) and the speed of sound for effectively one-dimensional waves along the BEC's long axis ($c_\text{s,1d} \approx 1.75$~mm s$^{-1}$).  
Utilizing numerical simulations below, we argue that this difference is due to a decrease in the effective hydrodynamic density and hence pressure jump across the shock front.

\subsection{Numerical Simulations and Vortex Turbulence}

\begin{figure}
\includegraphics[width=9cm]{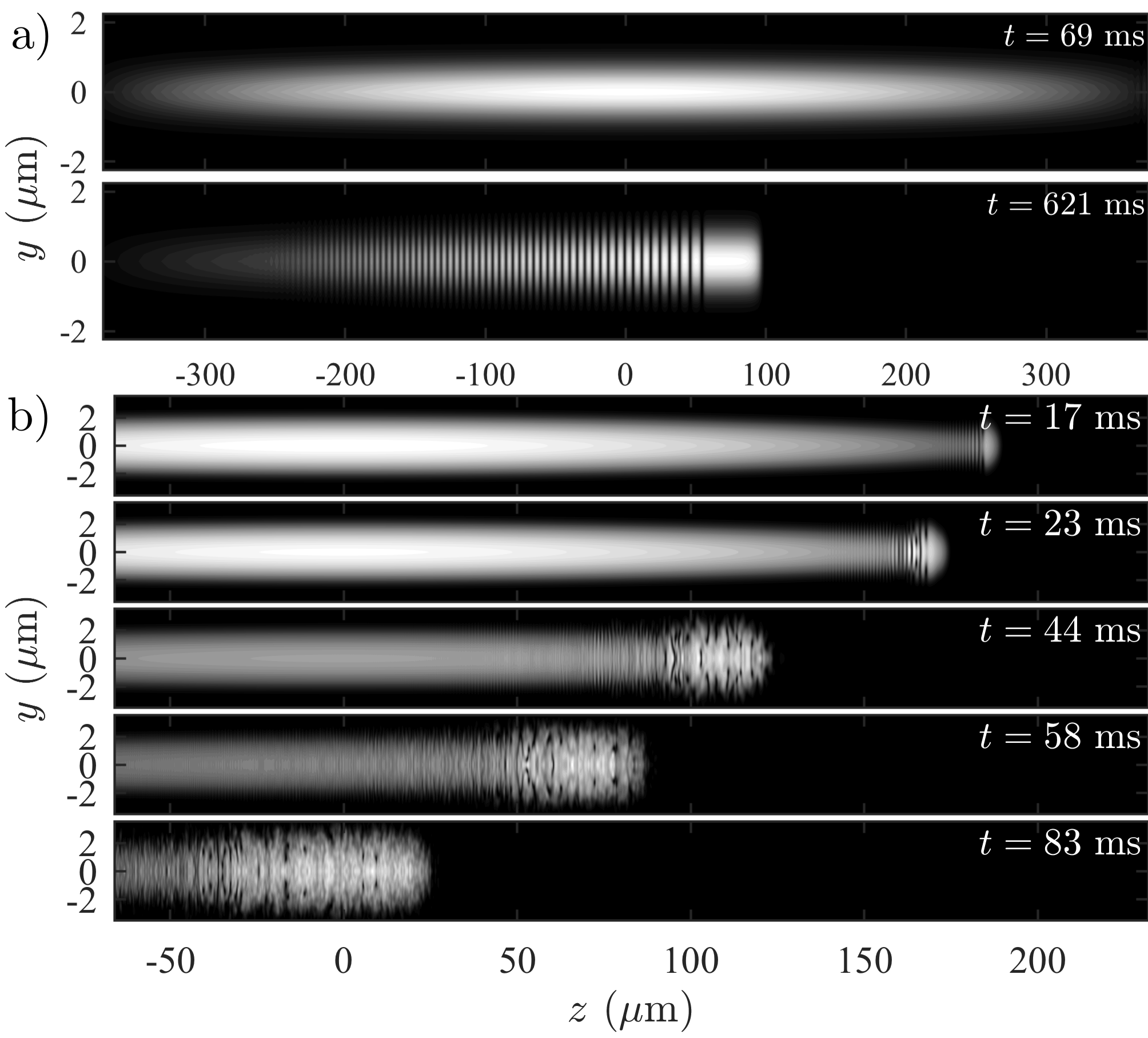}
\caption{\emph{Integrated cross sections of numerical simulations.}  {\bf a,}
  Quasi-1D dispersive shock wave in low atom number regime where
  $v_\text{p} = 0.41$~mm s$^{-1}$.  {\bf b,} Development of 3D turbulence and a
  viscous shock wave in the large atom number regime where $v_\text{p} = 2.44$~mm s$^{-1}$.}
\label{Fig:Zslices}
\end{figure}

The applicability of viscous shock theory may seem surprising for a nominally inviscid superfluid. 
Further insight can be gained from our comparative (3+1)D numerical simulations of the Gross-Pitaevskii equation, modeling the piston by a moving Gaussian potential that is swept through the BEC (see Methods and Supplementary Note 1 for details). 
While our numerics are performed in three spatial dimensions, the expected dimensionality of the dynamics can be classified in terms of the non-dimensional quantity (see \cite{Kevrekidis2015} and references therein)
\begin{equation}
  d = N \lambda \frac{a_\text{s}}{a_\text{ho}} ,
\end{equation}
where $N$ is the number of trapped atoms, $\lambda = \omega_{z}/\omega_{\perp}$ is the cylindrically symmetric trap aspect ratio, $a_\text{s}$ is the scattering length, and $a_\text{ho} = \sqrt{\hbar/m\omega_{\perp}}$ is the transverse harmonic trap length scale. 
When $d \ll 1$ the transverse dynamics are significantly constrained such that the BEC exhibits quasi-1D behavior.  
When $d \gg 1$, the BEC is no longer geometrically constrained and is truly described by 3D dynamics. 
A simulation in the 1D regime with corresponding dimensional spatial, temporal, and density scales of $L = a_\text{ho} = 0.713~\mu\text{m}$, $T = 1/\omega_{\perp} = 0.695~\text{ms}$, $\Gamma = 1/(4\pi a_\text{s} a_\text{ho}^2) = 29.1~\mu\text{m}^{-3}$, with $\omega_z = 2\pi\times 0.24~$Hz and $N = 8,\!732$ is shown in Fig.~\ref{Fig:Zslices}a with $d = 0.068$. 
A coherent soliton train or DSW is generated whose soliton edge speed quantitatively agrees with DSW theory \cite{Hoefer2008}, as shown in the Supplementary Fig.~2.  
For more details on this analysis, see Supplementary Note 2.

To match experimental parameters, the simulations in Fig.~\ref{Fig:Zslices}b consider a large number of atoms, $N = 410,\!005$, with a tighter trap geometry (a 10-fold increase in the longitudinal trap frequency) and the same scales $L$, $T$, and $\Gamma$ as in the low atom number case above. 
The dimensionality parameter is now $d \approx 32$, which places the dynamics well into the 3D regime.

\begin{figure} 
\includegraphics[clip=true,trim=14 0 0 0,width = 9.0cm]{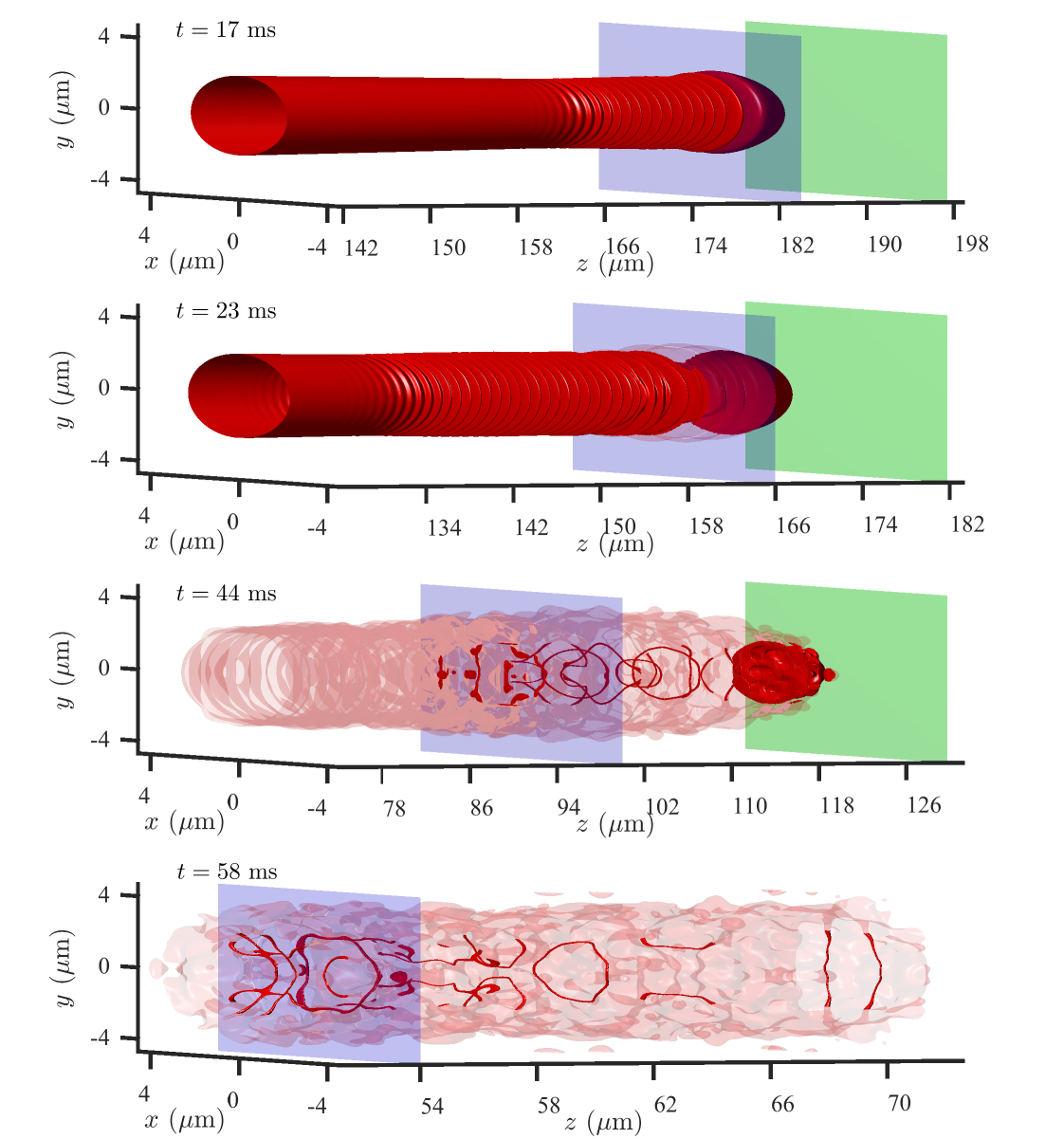}
\caption{\emph{Isosurfaces of 3-D Numerical Simulation.}  A piston
  front (rightmost, non-transparent green plane) sweeping into an
  elongated BEC at $v_\text{p} = 2.44$~mm s$^{-1}$. The shock front (leftmost, blue
  transparent plane) propagates through the BEC over time.  At
  time $t=58$~ms, only the shock front plane is shown.  In the
  plateau region between the two planes, vortex tangles are generated.
  The isosurface density value is $0.1\cdot \Gamma$, semi-transparent
  for $\sqrt{x^2+y^2} > 2.39~\mu\text{m}$ to visualize the BEC
  interior.}
\label{Fig:Isosurface}
\end{figure}

\begin{figure}
\includegraphics[width=9cm]{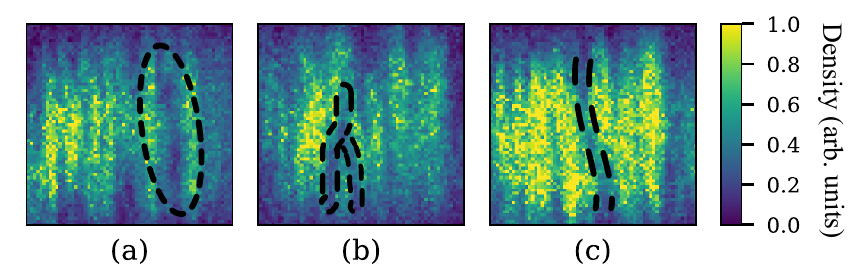}
\caption{\emph{Experimental evidence of vortex turbulence.} 
  The barrier is swept to the center of the BEC in 87~ms ($v_\text{p} = 2.5$~mm s$^{-1}$). 
  Within the plateau region, \textbf{a,} vortex rings, \textbf{b,} soliton y-forks, and \textbf{c,} soliton snaking are observed.
  Absorption images are taken after 10.1~ms of free expansion.
  Black dashed lines are intended as a guide to the eye.}
\label{Fig:ExpTurbulence}
\end{figure}

Isosurface plots corresponding to Fig.~\ref{Fig:Zslices}b panels are shown in Fig.~\ref{Fig:Isosurface}.  
Based on these simulations, the dynamics can be characterized as follows.  
At short times, a soliton train is initially formed, as would be expected in DSW theory (Fig.~\ref{Fig:Isosurface}, $t=17$~ms).  
Due to the large atom number--or, equivalently, weak transverse confinement--the soliton train rapidly undergoes a transverse, snake instability \cite{hoefer_onset_2016}, and vorticity emerges in the neighborhood of the shock front (Fig.~\ref{Fig:Isosurface}, $t=23$~ms, $t=44$~ms).
The snake instability manifests itself when the transverse harmonic oscillator length $a_\text{ho}$ exceeds the healing length by a factor of order one \cite{Muryshev2002}.  
As seen at $t=58$~ms and experimentally in Fig.~\ref{Fig:ExpTurbulence}, the plateau region hosts a variety of topological defects including vortex rings, lines, and vortex interactions.  
Experimental evidence for vorticity and soliton dynamics in the plateau regions obtained in absorption images after 10.1~ms expansion can be found in Fig.~\ref{Fig:ExpTurbulence}.  
Since the images in Fig.~\ref{Fig:ExpTurbulence} are integrated along the x-axis, vortex rings appear as two dark dots with a faint connection between them~\cite{Anderson2001}.  
Transformation of dark solitons into vortex rings has been argued to be responsible for an apparent inelasticity of collision events \cite{ricardo}. 
Furthermore, such rings have also been recently identified in bosonic \cite{bisset} and fermionic~\cite{zwierlein} systems.

\begin{figure}
  \centering
  \includegraphics[width=9cm]{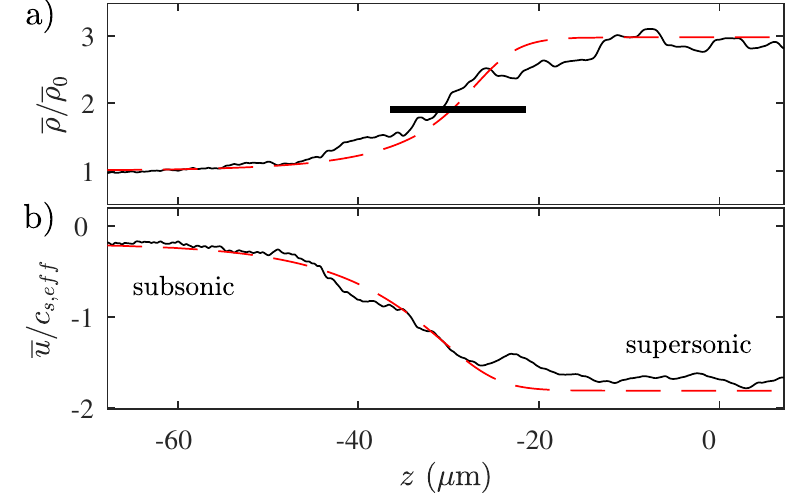}
  \caption{\emph{Shock profile.}  
  {\bf a,} Filtered density and {\bf b,} velocity profiles from 3D numerical simulation (black solid) in units of the downstream, subsonic flow density $\overline{\rho}_0$ and associated sound speed $c_\text{s,eff} = 1.35$ mm s$^{-1}$.  
  The red dashed curves correspond to an exact, viscous traveling wave solution of the piston problem for the 1D shallow water equations with an effective nondimensional viscosity parameter (see Methods) that reveals the shock structure and compares favorably to the experimentally measured shock width $15.5~\mu$m (horizontal segment).  
  This profile corresponds to Fig.~\ref{Fig:Zslices} at the time $t = 83$ ms.}
   \label{fig:vsw_fit}
\end{figure}

Both tight harmonic transverse confinement and quantum turbulence contribute to the highly inhomogeneous background through which the shock propagates.  
The transverse trap width $a_\text{ho}$ and the healing length in the condensate center are less than $1~\mu$m, thus these characteristic length scales of the system  are significantly smaller than the measured shock width $\Delta x_\text{w} \approx 15.5~\mu$m. 
Consequently, we argue that the shock experiences an effectively averaged or filtered turbulent field and an associated decrease in the effective density (and hence pressure).
To quantify this, we use a technique from the large eddy simulation (LES) framework \cite{Garnier2009} to filter by convolving the hydrodynamic field from the simulation in Fig.~\ref{Fig:Zslices} at $t = 83$~ms.
The filtering is done with a kernel, $K(\mathbf{r}) = w^{-3}$, for $\mathbf{r}$ in a cube with side length $w$ centered at the origin, and zero otherwise. 
This leads to a reduction in the central, filtered condensate density, $\overline{\rho}(\mathbf{r},t) = (K * \rho)(\mathbf{r},t) = \int K(\mathbf{r} - \mathbf{r}') \rho(\mathbf{r}',t)\, d\mathbf{r}'$, when compared to the unfiltered, initial, unperturbed density.
By choosing a cube side length of $w = 5.35~\mu$m, we obtain excellent agreement with our observed reduced speed of sound, $c_\text{s,eff}$, independently fit in Fig.~\ref{Fig:MainResults}.
This kernel size leads to a 3.35-fold reduction in the central, filtered condensate density.
In this case, the effective sound speed satisfies $c_\text{s,eff} = \sqrt{4\pi \hbar^2 a_\text{s} \overline{\rho}/m^2} = c_\text{s,bulk}/\sqrt{3.35} \approx 1.35$ mm s$^{-1}$. 
The filtered hydrodynamic field is well-described by an exact viscous shock profile of the viscous shallow water equations as is shown in Fig.~\ref{fig:vsw_fit} where $\overline{u} = (K * \rho \mathbf{u})/\overline{\rho}$ is the Favre-averaged velocity \cite{Garnier2009} (see Methods).

Support for this interpretation can be found in the literature on shocks in turbulent gases \cite{Garnier2009,tanaka}. 
Turbulent dynamics modeled as dissipation at larger scales is the basis for the effective eddy viscosity in LES \cite{Garnier2009}. 
We view the quantum turbulence here in a similar fashion. In a coarse-grained description, the vortex excitations act as a reservoir for energy dissipated from the large amplitude wave excitations generated at the large-scale shock front.
We compare an exact viscous shock profile of the viscous shallow water equations with our filtered 3D numerical simulations in Fig.~\ref{fig:vsw_fit}. 
This reveals a shock structure consisting of a smooth transition from a nonzero subsonic flow ($|\overline{u}| < c_\text{s,eff}$) ahead of the shock to supersonic flow ($|\overline{u}| = v_\text{p} > c_\text{s,eff}$) behind the shock. 
The width of this transition is proportional to the effective dissipation experienced by the shock. 
The numerically observed nonzero mean flow ahead of and in the same direction as shock propagation is additional evidence for a decrease in the shock's pressure (hence density) as observed in numerical simulations of shock waves in a turbulent gas \cite{tanaka}.
We stress that the only fitting parameters in Fig.~\ref{fig:vsw_fit} are the filtering length scale, the shock width, and the mean flow ahead of the shock. 
All remaining quantities---mean density/velocity behind the shock, the shock speed, and the viscous shock profile---are completely determined by the Rankine-Hugoniot jump conditions for a piston in a viscous fluid. 
The quantitative agreement between viscous shock theory, the filtered, conservative BEC simulation, and experiment (Fig.~\ref{Fig:MainResults}) constitutes strong support for our conclusions. 
For additional analysis of the turbulence and associated energy production, see Supplementary Note 3 along with corresponding Supplementary Fig.~3 and 4.

\subsection{Rarefaction Waves}

The emergence of a shock front as described above crucially depends on the presence of a finite background density through which the front propagates. 
In the absence of such a background density, the phenomenology is completely different and rarefaction waves emerge (Fig.~\ref{Fig:SweepSequence}(b), 140~ms). 
Rarefaction waves, which are commonly discussed in the context of the shock tube problem in gas dynamics \cite{Liepmann1957} or the dam break problem in shallow water \cite{Leveque2002}, form when a region of high density expands into a region of zero density (vacuum).

For dilute-gas BECs, we can study the formation of rarefaction waves in a clean, unperturbed setting by starting with $4.1\times10^5$~atoms confined in the left half of the harmonic trap. 
A repulsive barrier initially prevents the atoms from spreading out into the empty right half of the trap. 
When the barrier is suddenly removed, the BEC begins to spread out to the right and the front edge of the BEC density assumes a parabolic shape. 
For further details and experimental cross sections, see Supplementary Note 4. 
The self-similar expansion of the rarefaction front is in stark contrast to the dynamics of a sharp shock front. 
The parabolic shape is consistent with the predictions of these types of waves in a Bose-condensed gas \cite{Hoefer2006,Gurevich1995,Hoefer2009,El1995}. 
The theory also predicts that the expanding edge propagates at twice the local speed of sound \cite{Hoefer2009,El1995}.

\begin{figure}
\includegraphics[width=9cm]{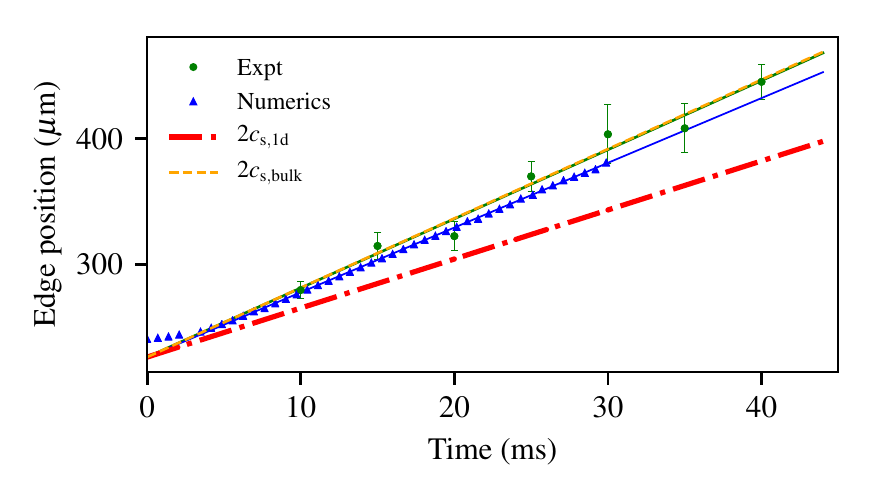}
\caption{\emph{Rarefaction waves.}  
A BEC initially confined to the left half of the trap is suddenly allowed to spread to the right. 
The plot shows the edge position vs. time, where experiment (green dots) and numerics (blue triangles) are plotted overlaid with expected results for $2c_\text{s,bulk}$ (orange dashed) and $2c_\text{s,1d}$ (red dot-dashed).  
Experimental data are mean $\pm$ s.d. for 5 runs at each measured time, where the green and blue solid lines are best fits to experiment and numerics, respectively. 
See text for more information.}
\label{Fig:Rarefaction}
\end{figure}

We test this prediction in experiment by fitting the parabolic expanding edge of the BEC integrated cross section at various times during the expansion, and deduce from this the edge propagation speed (See Fig.~\ref{Fig:Rarefaction} and Supplementary Fig.~5).  
The edge speed determined from our experiment ($5.50 \pm 0.33$~mm s$^{-1}$) and from matching numerics ($5.16 \pm 0.01$~mm s$^{-1}$) is in good agreement with the predicted behavior based on the 3D speed of sound (i.e., $2\times c_\text{s,bulk} = 5.6$~mm s$^{-1}$).

The data and numerics in Fig.~\ref{Fig:Rarefaction} are also compared to $2\times c_\text{s,bulk}/\sqrt{2}=2\times c_\text{s,1d}=4.02$~mm s$^{-1}$, which is the expected rarefaction edge propagation speed in a 1D channel ($d \ll 1$).  
We see a clear deviation from this behavior, further indicating the fully 3-dimensional structure of our system ($d \gg 1$) and the inapplicability of 1-dimensional DSW theory.

\section{Discussion}

In conclusion, we have observed and analyzed intriguingly rich dynamics in a quantum mechanical piston shock.  
For our typical experimental parameters, the dynamics are described by dissipative, rather than dispersive, shock waves.  
The piston provides both the source of the shock front and the generation of superfluid quantum turbulence, manifested through the development of vortical patterns and dispersive waves, via a transverse, snake instability of a planar soliton train.  
We argue that an effective dissipation arises in a nominally inviscid superfluid as a consequence of the dissipation of large amplitude excitations from the large scale shock front into small-scale vortex excitations.

Our experiments provide a versatile platform for the investigation of quantum turbulence, which is currently an area of intense research efforts in both cold atom and superfluid helium systems. 
Further studies into lower dimensional systems with similar geometries may also be of interest to determine the effect of dimensionality on quantum turbulence.

\section{Methods}

\subsection{Experimental Procedure}
Our experimental setting consists of an elongated BEC of $4.1\times10^5$ $^{87}$Rb atoms confined in an optical dipole trap with trap frequencies $\{\omega_x, \omega_y, \omega_z\} = 2\pi \times \{229, 222, 2.4\}$~Hz.
We estimate the atom number in the BEC by fitting integrated cross sections of the absorption images to the numerical ground state.  
Temperature is estimated to be $\ll T_\text{c}$, the critical temperature of the BEC, given no observable thermal cloud.  
We note that for velocities of $4$~mm s$^{-1}$ and lower, we see no noticeable atom loss in experiments.  
Agreement between experiment and zero-temperature GP numerics throughout the complete barrier sweeps further corroborates this point.  

The piston is generated by a repulsive laser beam of wavelength $\lambda_\text{piston} = 660$~nm and an elliptical cross section of Gaussian waists $\{w_x, w_z\} \approx \{68, 11\}~\mu$m.  
The barrier is swept from right to left (negative $z$-direction) using a high-speed mirror galvanometer. 
Effects of the initial acceleration of the galvanometer are avoided by initializing the barrier sweep far from the right edge of the BEC.  
This is done such that the acceleration range occurs outside of the BEC. Absorption images are taken after 2~ms expansion time to avoid image saturation.

\subsection{Three-dimensional simulations}
\label{sec:three-dimens-simul}

We perform (3+1)D numerical simulations of the Gross-Pitaevskii equation \cite{Kevrekidis2015}.  
The initial condition consists of the ground state solution in the form $\psi(\mathbf{r},t) = f(\mathbf{r}) e^{-i\mu t}$ in the presence of the harmonic trap without the barrier.  
$\mu$ is the chemical potential, determined by the total number of atoms $N$ in the condensate according to
\begin{equation}
  \label{eq:6}
  \int_{\R^3} f^2(\mathbf{r}) \, \mathrm{d} \mathbf{r} =
  N\Gamma^2a_\text{ho}^3 ,
\end{equation}
where $a_\text{ho} = \sqrt{\hbar/(m\omega_\perp)}$ and $\Gamma = 1/(4\pi a_s a_{ho}^2) = 29.1~\mu\text{m}^{-3}$.  
We utilize a Fourier spatial discretization and a second order split-step method, exactly integrating the linear and nonlinear/potential terms separately.  
For the simulation of the experiment, we use a grid spacing of $0.071~\mu\text{m}$ on a box of size $8.8\times 8.8\times 464~\mu\text{m}^3$ with a time-step of $0.0013$~ms.  
The simulation in Figs.~\ref{Fig:Zslices}b and \ref{Fig:Isosurface} exhibits a healing length of $0.21~\mu$m at the initial trap center and $0.16~\mu$m in the plateau region.  
The simulation in Fig.~\ref{Fig:Zslices}a exhibits a healing length of $1.24~\mu$m at the initial trap center and $0.76~\mu$m in the DSW plateau region.

\subsection{Viscous Shock Fitting}
\label{sec:visc-shock-fitt}

Following Favre averaging \cite{Garnier2009}, filtering is performed by convolving the 3D density and momentum in the longitudinal direction with a cube of side length 5.35 $\mu$m.  
This cube size results in the downstream speed of sound $c_\text{s,eff} = 1.35$ mm s$^{-1}$, used in the main text to explain the experimental and numerical observations.  
This cube size is a plausible filter in the context of large eddy simulation modeling as it is an intermediate length scale to the measured shock width (15.5~$\mu$m) and the healing length ($\sim 0.2~\mu$m).  
The velocity is recovered by dividing the filtered momentum by the filtered density (Favre filtering).  
The shock profile is obtained from an exact, traveling wave solution of the one-dimensional shallow water equations (the dispersionless limit of the Gross-Pitaevskii equation without a potential) with an additional phenomenological viscous term
\begin{equation}
  \label{eq:5}
  \begin{split}
    \overline{\rho}_t + (\overline{\rho} \overline{u})_z &= 0 \\
    (\overline{\rho} \overline{u})_t + (\overline{\rho} \overline{u}^2
    + \frac{1}{2} \overline{\rho}^2)_z &= \nu \overline{u}_{zz} .
  \end{split}
\end{equation}
for the nondimensional, filtered density $\overline{\rho}$, velocity $\overline{u}$, and viscosity parameter $\nu > 0$. 
The traveling wave speed, plateau (rightmost) density and velocity are obtained from the Rankine-Hugoniot viscous shock conditions for a piston.  
The fitting parameters are the nondimensional viscosity $\nu = 25$ and the traveling wave center $z_0 = -25~\mu$m, obtained by minimizing the sum of the absolute differences between the filtered and traveling wave superfluid velocity.

\noindent \textbf{Code availability}
All relevant code used for numerical studies in this work is available from the corresponding authors on reasonable request.

\noindent \textbf{Data availability} 
All relevant experimental and numerical datasets in this work are available from the corresponding authors on reasonable request.

\noindent \textbf{Acknowledgements} 
M.E.M. and P.E. acknowledge funding from the
National Science Foundation (NSF) through Grant No. PHY-1306662 and PHY-1607495. 
M.A.H is partially supported by NSF CAREER DMS-1255422.
K.J. is supported by National Science Foundation under award numbers DMS1317666 and EAR-1620649.
P.G.~Kevrekidis is partially supported by NSF-PHY-1602994.

\noindent \textbf{Author contributions} 
M.E.M, P.E. and M.A.H
conceived the experiment and theoretical modeling; 
M.E.M. and P.E. performed the experiments; 
M.A.H performed the theoretical
calculations, consulting with P.G.K. and K.J.; 
P.E. supervised the project. All authors discussed the
results and contributed to the writing of the manuscript.

\noindent \textbf{Competing interests}: The authors declare no
competing financial or non-financial interests.\newline


\end{document}


\section{Supplementary Material}

\renewcommand{\theequation}{\arabic{equation}}
\setcounter{equation}{0}
\renewcommand{\figurename}{Supplementary Figure}
\setcounter{figure}{0}

\subsection{Supplementary Note 1}
\label{sec:governing-equation}

The Gross-Pitaevskii equation
\begin{equation}
  \label{eq:1}
  i\psi_t = - \frac{1}{2} \nabla^2 \psi + V \psi + |\psi|^2 \psi,
  \quad \mathbf{r} \in \R^3, \quad t > 0
\end{equation}
is used to model the quantum piston problem whereby a BEC is condensed
to its ground state in a cigar-shaped harmonic trap
\begin{align}
  \label{eq:2}
  V_{\rm ho}(x,y,z) &= \frac{1}{2}((\lambda_x x)^2 + y^2 +(\lambda_z
  z)^2 ), \\
  \lambda_x &= 0.969 \approx 1, \quad \lambda_z = 0.01\ll 1 .
\end{align}
The long axis of the BEC is along the $z$-axis.  The piston is a repulsive laser sheet that is swept across the trapped BEC, modeled by a moving Gaussian potential
\begin{align}
  \label{eq:3}
  V_{\rm p}(x,y,z) &= V_0 \exp \left [-\frac{x^2+y^2}{2s_\rho^2} -
    \frac{z^2}{2 s_z^2}  \right ], \\
    V_0 &= 155, \quad s_{\rho} = 95.3, \quad s_z = 15.4 .
\end{align}
The potential in eq.~(\ref{eq:1}) consists of the piston translated axially and the stationary trap
\begin{equation}
  \label{eq:4}
  V(x,y,z,t) = V_{\rm ho}(x,y,z) + V_{\rm p}(x,y,z + v_{\rm p}t-z_0) ,
\end{equation}
with piston velocity $v_{\rm p}$ and the initial piston location
$z_0 = 365$.

These are dimensionless equations.  
The corresponding dimensional spatial, temporal, and density scales are $a_y = \sqrt{\hbar/(m \omega_y)}$, $T = 1/\omega_y$, and $\Gamma = 1/\sqrt{4\pi a_{\rm s} a_y^2}$, respectively.


Equation~\eqref{eq:1} can be written in hydrodynamic form with the
exact transformation $\psi = \sqrt{n} e^{i\phi}$,
$\mathbf{u} = \nabla \phi$
\begin{equation}
  \label{eq:11}
  \begin{split}
    n_t + \nabla\cdot(n\mathbf{u}) &= 0 \\
    (n \mathbf{u})_t + \nabla \cdot (n \mathbf{u} \otimes \mathbf{u}) +
    \frac{1}{2} \nabla n^2 &= \frac{1}{4} \nabla ( n \nabla \otimes
    \nabla \log n ) - \nabla V.
  \end{split}
\end{equation}
Restricting to one-dimensional, planar shocks
$u = \mathbf{u} \cdot \mathbf{\hat{z}}$
($\mathbf{\hat{z}} = (0,0,1)^T$), neglecting the potential $V$ and the
dispersive term $[n(\log n)_{zz}]_z$ yields the long-wave hydrodynamic
equations
\begin{align}
  \label{eq:13}
  n_t + (n u)_z &= 0 \\
  \label{eq:14}
  (nu)_t + (n u^2 + \frac{1}{2} n^2)_z &= 0 .
\end{align}
These equations are equivalent to the shallow water equations
\cite{Leveque2002}.  We now solve the viscous shock problem by
invoking the Rankine-Hugoniot jump relation \cite{Leveque2002}
\begin{align}
  \label{eq:15}
  -s [n] + [nu] &=0 \\
  \label{eq:16}
  -s [nu] + [nu^2 + \frac{1}{2} n^2] &= 0,
\end{align}
where $[\cdot ]$ represents a jump in the quantity across the
discontinuous shock front and $s$ is the shock speed.  If we normalize
to quiescent downstream conditions $n_+ = 1$, $u_+ = 0$, then we can
solve Eqs.~\eqref{eq:15} and \eqref{eq:16} for the jump in density and
velocity $n_- > n_+$, $u_- > u_+$ to give a shock satisfying
\begin{align}
  \label{eq:17}
  u_- = \frac{n_- - 1}{\sqrt{2}}\sqrt{1/n_-+1}, \quad s = \frac{n_-
  u_-}{n_- - 1} .
\end{align}
The piston problem is solved by equating the flow speed to the piston
speed $u_- = v_p$.  The equations in \eqref{eq:17} are used to obtain
the theoretical viscous shock speed $s$ and plateau density $n_-$
curves in Fig.~2 of the main manuscript where the speed is normalized by
the effective speed of sound $c_\text{s,eff} = 1.35$ mm s$^{-1}$ and the density
is normalized by the maximum density at the BEC center.

We note that viscous effects lead to the imposition of the jump
conditions \eqref{eq:15}, \eqref{eq:16} for the viscous shock.
However, the magnitude of the viscosity does not influence these
conditions.  The primary role of the viscosity is to determine the
shock structure, i.e., the width over which the flow transitions from
$(n_-,u_-)$ to $(n_+,u_+)$.  

\subsection{Supplementary Note 2}
\label{sec:pist-disp-shock}

For completeness, we quote the formulas in normalized units of the
sound speed $c_s$ and density for a dispersive shock wave (DSW)
solution, i.e., a shock wave in which viscosity is negligible relative
to dispersion \cite{Hoefer2008}.  In contrast to viscous shock waves,
dispersive shock waves exhibit two speeds of propagation, the
trailing, large amplitude soliton edge and the leading, small
amplitude harmonic edge.  For a piston moving with speed $v_p$ into a
quiescent BEC with unit density and zero velocity, the plateau density
$n_- = (v_\text{p}/2 + 1)^2$ saturates when $v_\text{p} = 2$, twice the speed of
sound in the quiescent BEC.  For piston speeds larger than $v_\text{p} = 2$,
the dispersive shock wave is oscillatory up to the piston and the peak
normalized density oscillation is $4$.  The dispersive shock wave
soliton edge moves with speed
\begin{equation*}
  v_\text{soli} =
  \begin{cases}
    v_\text{p}/2+1 & 0 < v_\text{p} \le 2, \\
    v_\text{p} - 2 \left [ 1 - \frac{v_\text{p} E(4/v_\text{p}^2)}{(v_\text{p}-2)K(4/v_\text{p}^2)}
    \right ]^{-1} & v_\text{p} > 2 ,
  \end{cases}
\end{equation*}
where $K$ and $E$ are the complete elliptic integrals of the first and
second kind.  Note that there is an error in \cite{Hoefer2008} for the
dispersive shock wave large amplitude edge speed when $v_\text{p} > 2$.  The
dispersive shock wave harmonic edge moves with speed $v_\text{har} =
(2v_\text{p}^2+4v_\text{p}+1)/(v_\text{p}+1)$ for all piston speeds $v_\text{p} > 0$.  The best
fit of the experimental and numerical simulation data is shown in
Supplementary Fig.~\ref{fig:dsw_speed_fit} using the speed of sound as the sole
fitting parameter (as in the viscous shock case reported in Fig.~2 of
the main manuscript).  Here, the best fit effective speed of sound is
$c_\text{s,eff} = 1.55$ mm/s, but the data does not agree with both
the shock speed and plateau density simultaneously across the range of
piston speeds surveyed.
\begin{figure}
  \centering
  \includegraphics[width=9cm]{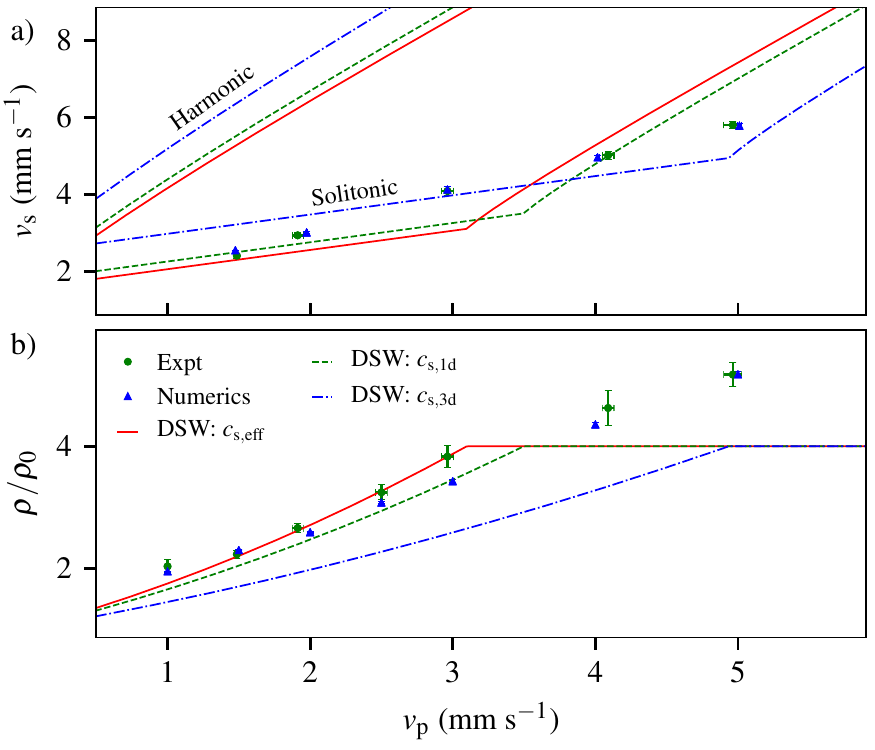}
  \caption{{\em Shock Speed, Peak Height vs. Piston Speed.}
  Best fit of experimental (green dots) and numerical (blue triangles) data to the
    piston DSW closure relations with $c_\text{s,eff} = 1.55$ mm/s, 
    $c_\text{s,1d}=1.75~$mm/s, and $c_\text{s,3d} = 2.47~$mm/s.
    \textbf{a,} Both the DSW harmonic edge
    speed, $v_\text{har}$ (upper curves), and soliton
    edge speed, $v_\text{soli}$
    (lower curves), are plotted. For the detailed definition of these quantities,
    see Supplementary Note 2.
    \textbf{b,} The maximum normalized peak height with respect to DSW theory.
    For DSW theory, the saturation of the plateau height and a 
    discontinuity of $v_\text{soli}$ occur when $v_\text{p} \ge 2 c_\text{s}$.
    Weighted mean $\pm$ s.d. are plotted for three (a) and five (b) data runs.
    }
  \label{fig:dsw_speed_fit}
\end{figure}

\begin{figure}
  \centering
  \includegraphics[width=9cm]{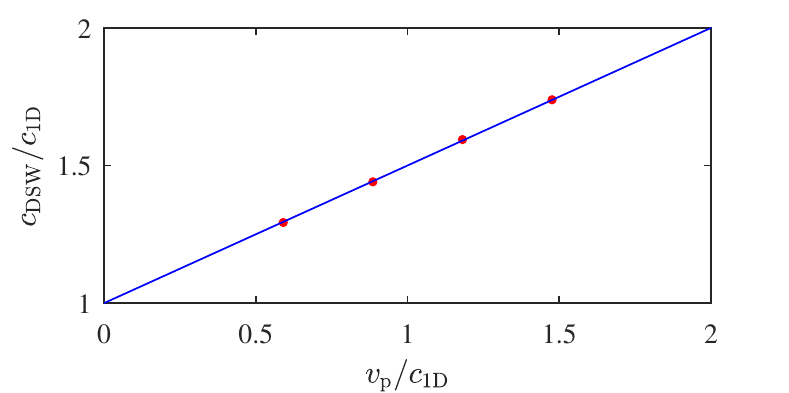}
  \caption{{\em DSW trailing edge soliton speed vs. piston speed.} 
  Comparison between 1D theory (circles) and 3D numerics
    (solid line).}
\label{fig:dsw_theory_numerics}
\end{figure}

We have verified the predictions of DSW theory by 3D numerical
simulations with sufficiently tight confinement ($\lambda_x = 0.969$,
$\lambda_z = 0.0010$, number of atoms $N = 8732$).  In this case, the
dimensionality parameter $d = 0.068$ is much smaller than unity.
Additionally,
\begin{equation*}
\left ( \frac{N a_{\rm s}}{\sqrt{\lambda_z} a_y} \right )^{1/3} = 12.6,
\end{equation*} 
which is much larger than unity and therefore the ground state
can be approximately factored into the Thomas-Fermi (dispersionless)
approximation for the long ($z$) direction and the harmonic oscillator
ground state for the transverse ($x$-$y$) directions \cite{Mateo2008, Menotti2002}.  This enables the
approximate solution of the 3D BEC piston problem using the 1D DSW
results.  Supplementary Fig.~\ref{fig:dsw_theory_numerics} shows a convincing
comparison between the DSW soliton edge speed $c_{\rm DSW}$ from 3D
numerical simulations and the 1D theory for a range of piston speeds.
The effective speed of sound is the usual 1D speed of sound
$\sqrt{n_0/2}$ where $n_0$ is the peak density at the trap center,
i.e., no fitting was used in Supplementary Fig.~\ref{fig:dsw_theory_numerics}.

\subsection{Supplementary Note 3}
\label{sec:energetics}

\begin{figure}
  \centering
  \includegraphics[width=18cm,clip=true,trim=35 0 50 0]{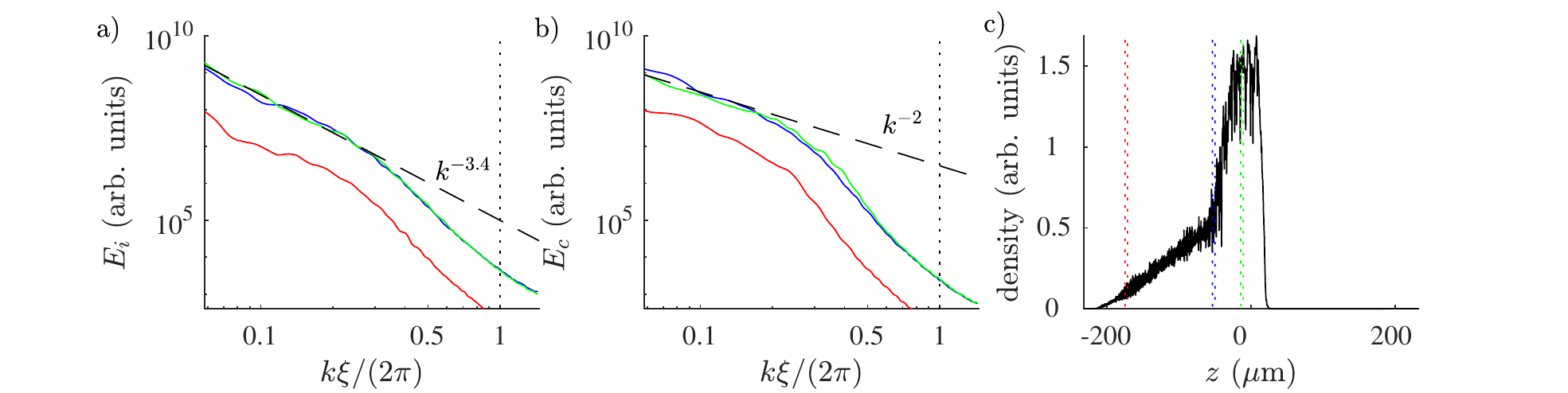}
  \caption{{\em Incompressible and compressible kinetic energy spectra.}
  {\bf a,} Azimuthally averaged, 2D incompressible and
  {\bf b,} compressible kinetic energy spectra obtained from the
    average of multiple spatial slices taken transverse to the piston
    direction at the locations identified by correspondingly colored vertical dashed lines
    in {\bf c}.  The spectrum is normalized to the healing length at the
    trap center for the equilibrium condensate ($0.21~\mu$m) taken
    prior to piston motion.  The incompressible and compressible
    spectrum exhibit power law decay $\sim k^{-3.4}$ and
    $\sim k^{-2}$, respectively over a range of wavenumbers.}
  \label{fig:spectra}
\end{figure}

Supplementary Fig.~\ref{fig:spectra}a and b reports 2D incompressible and compressible kinetic energy spectra, respectively, extracted from three regions (red, blue and green) in the flow indicated in Supplementary Fig.~\ref{fig:spectra}c.  
The two rightmost intervals (blue and green) exhibit similar spectral features and a fitted $k^{-3.4}$ scaling.  
The reported energy spectra are the $z$-averaged spectra computed from spatial slices of the density-scaled velocity field $\sqrt{\rho} \mathbf{u} = \mathbf{v}_\text{i} + \mathbf{v}_\text{c}$, decomposed into its incompressible $\mathbf{v}_\text{i}$ (divergence free) and compressible $\mathbf{v}_\text{c}$ (curl free) components
\begin{equation}
  \label{eq:5}
  E_\text{i,c}(k) = \int_{a}^b \int_{0}^{2\pi} |\hat{\mathbf{v}}_\text{i,c}(k \cos
  \varphi,k\sin\varphi,z)|^2\,d\varphi dz,
\end{equation}
where 
\begin{equation}
  \label{eq:6}
  \hat{\mathbf{v}}_\text{i,c}(k_x,k_y,z) = \int_{\mathbb{R}^2}
  \mathbf{v}_\text{i,c}(x,y,z) e^{-i(k_x x + k_y y)}\, dx dy
\end{equation}
is the two-dimensional Fourier transformation in the transverse
$x$-$y$ directions.

\begin{figure}
  \centering
 \includegraphics[width=9cm]{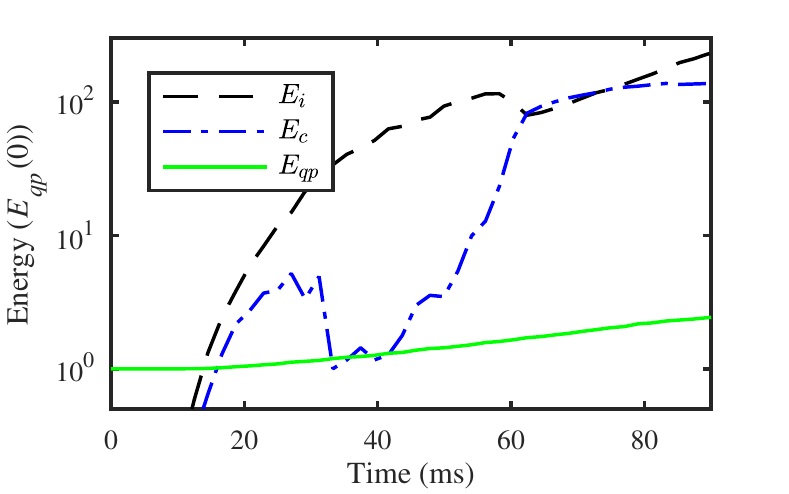}
 \caption{\emph{Evolution of Energies.}
 Total quantum pressure (green solid), incompressible
     (black dashed) and compressible (blue dot-dashed) kinetic energies from
     simulation in Fig.~3b and Fig.~4 of the main manuscript.
   The decrease in compressible kinetic energy at $t \approx 30$ ms
   coincides with the transverse (snake) instability and subsequent
   break-up of the transient DSW.  From $\approx 40$ ms onwards, the
   total quantum pressure, incompressible, and compressible energies
   increase due to the continual production of quantized vorticity and
   dispersive waves.  All energies are normalized by the initial
   quantum pressure energy.}
  \label{fig:energies}
\end{figure}

Supplementary Fig.~\ref{fig:energies} reports the temporal development of the
total quantum pressure energy
$E_\text{qp}(t) = \frac{1}{2} \int |\nabla \sqrt{n}|^2\,
\mathrm{d}\mathbf{r}$, total incompressible
$E_\text{i} = \frac{1}{2} \int |\mathbf{v}_\text{i}|^2\,\mathrm{d}\mathbf{r}$ and
compressible
$E_\text{c} = \frac{1}{2} \int |\mathbf{v}_\text{c}|^2\,\mathrm{d}\mathbf{r}$
kinetic energies where $\mathbf{v}_\text{i}$ and $\mathbf{v}_\text{c}$ are the
divergence free and curl free components, respectively, of the scaled
velocity field $\mathbf{v} = \sqrt{n} \mathbf{u}$.  The quantum
pressure energy highlights regions of the flow exhibiting large
density gradients such as those that occur in solitons and vortex
lines.  Consequently, it scales with the vortex line length and
therefore represents a measure of the amount of quantized vorticity.
In addition to the main text's reported vortex structures, the
monotonic increase of $E_\text{qp}$ as a function of time is an additional
indication of continued turbulence production as the piston is swept
through the BEC.

\subsection{Supplementary Note 4}
\label{sec:rarefactionsupplement}

To experimentally study rarefaction waves, we adiabatically sweep the barrier from the right (outside of the BEC) to the center of the BEC such that no excitations are formed during the sweep.  After this sweep, all atoms are confined in the left half of the trap.  The background density in the area to the right of the BEC is determined to be negligible.  The barrier is then jumped off and images are subsequently taken at $t = \{10,15, 20, 25, 30, 35, 40\}$~ms.  A direct comparison to numerics can be found in Supplementary Fig.~\ref{Fig:RarefactionSeq}.  

\begin{figure}
\includegraphics[width=9cm]{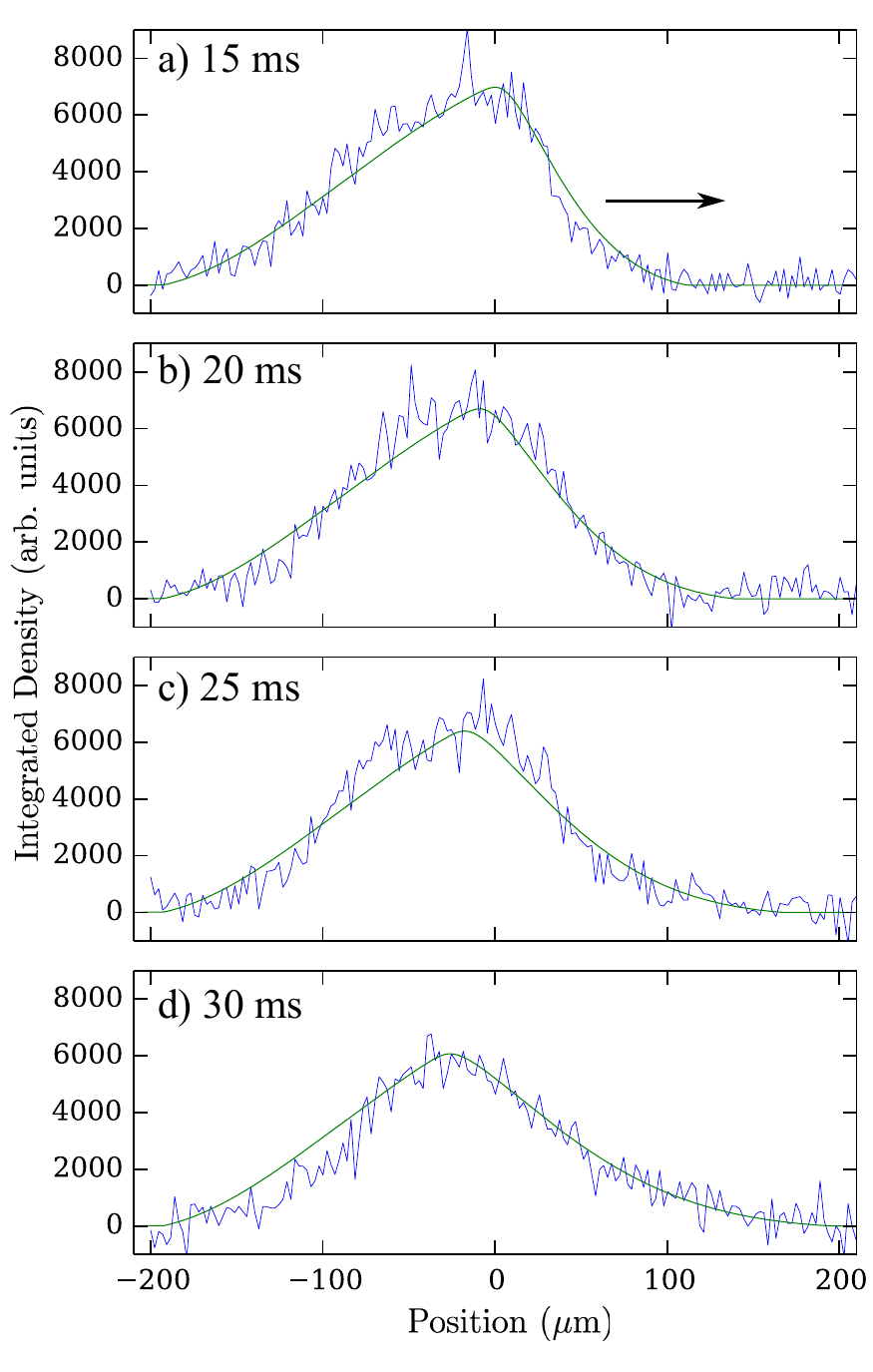}
\caption{\emph{Integrated cross section of rarefaction waves.}  
Cross sections for experiment (blue) and numerics (green) are plotted for {\bf a,} 15~ms, {\bf b,} 20~ms, {\bf c,} 25~ms, {\bf d,} 30~ms, after the barrier has been jumped off.  }
\label{Fig:RarefactionSeq}
\end{figure}

To analyze the data and corresponding numerics found in the main text, we first make an integrated cross section of the data as seen in Supplementary Fig.~\ref{Fig:RarefactionSeq}, and take the square root of this cross section. The right half of the cross section is now linear, as expected from a quadratic profile, and is fitted with a linear trend line.  The fit is extended to the point where it intersects zero, which is recorded as the edge of the rarefaction wave.  Results of this analysis can be found in Fig.~6 of the main manuscript.
